\documentclass[twocolumn]{aastex63}
\usepackage{amsmath}
\usepackage{color}
\usepackage{graphicx}
\usepackage{textcomp}
\usepackage[T1]{fontenc}

\shorttitle{Anisotropic multimessenger signals from merger NDAFs}
\shortauthors{Qi et al.}

\begin{document}

\title{Anisotropic multimessenger signals from black hole neutrino-dominated accretion flows with outflows in binary compact object mergers}

\correspondingauthor{Tong Liu}
\email{tongliu@xmu.edu.cn}

\author{Yan-Qing Qi}
\affiliation{Department of Astronomy, Xiamen University, Xiamen, Fujian 361005, China}

\author[0000-0001-8678-6291]{Tong Liu}
\affiliation{Department of Astronomy, Xiamen University, Xiamen, Fujian 361005, China}
\affiliation{SHAO-XMU Joint Center for Astrophysics, Xiamen University, Xiamen, Fujian 361005, China}

\author{Bao-Quan Huang}
\affiliation{Department of Astronomy, Xiamen University, Xiamen, Fujian 361005, China}

\author{Yun-Feng Wei}
\affiliation{Department of Astronomy, Xiamen University, Xiamen, Fujian 361005, China}

\author{De-Fu Bu}
\affiliation{Key Laboratory for Research in Galaxies and Cosmology, Shanghai Astronomical Observatory, Chinese Academy of Sciences, Shanghai 200030, China}
\affiliation{SHAO-XMU Joint Center for Astrophysics, Xiamen University, Xiamen, Fujian 361005, China}

\begin{abstract}
A black hole (BH) hyperaccretion system might be born after the merger of a BH and a neutron star (NS) or a binary NS (BNS). In the case of a high mass accretion rate, the hyperaccretion disk is in a state of neutrino-dominated accretion flow (NDAF) and emits numerous anisotropic MeV neutrinos. Only a small fraction of these neutrinos annihilates in the space outside of the disk and then launch ultrarelativistic jets that break away from the merger ejecta to power gamma-ray bursts. Mergers and their remnants are generally considered sources of gravitational waves (GWs), neutrinos, and kilonovae. Anisotropic neutrino emission and anisotropic high-velocity material outflows from central BH-NDAF systems can also trigger strong GWs and luminous disk-outflow-driven (DOD) kilonovae, respectively. In this paper, the anisotropic multimessenger signals from NDAFs with outflows, including DOD kilonovae, MeV neutrinos, and GWs, are presented. As the results, the typical AB magnitude of the DOD kilonovae is lower than that of AT 2017gfo at the same distance, and it decreases with increasing viewing angles and its anisotropy is not sensitive to the outflow mass distribution but mainly determined by the velocity distribution. Since neutrinos with $\gtrsim 10~\rm MeV$ are mainly produced in the inner region of the disk, they will be dramatically deflected to a large viewing angle by relativity effects. Moreover, the strains of GWs induced by anisotropic neutrinos increase with increasing viewing angles. The accumulation of multimessenger detection of the BH-NS/BNS mergers with different viewing angles might further verify the existence of NDAFs with outflows.
\end{abstract}


\section{Introduction}

In the era of multimessenger astronomy, the merger of a black hole (BH) and a neutron star (NS) or a binary NS (BNS) offers four mainly different probes, i.e., gamma-ray bursts (GRBs), kilonovae, neutrinos, and gravitational waves (GWs). As the binaries approach each other during the merger, some of their mass becomes tidally disrupted, forming a central compact object, which may be a magnetar \citep[e.g.,][]{Duncan1992,Usov1992,Zhang2001,Dai2006} or a BH \citep[e.g.,][]{Woosley1993,Lei2009,Lei2013,Liu2017a}. The inspiral and coalescence of compact objects are prime sources of GWs \citep[e.g.,][]{Schutz1989,Cutler1994,Sathyaprakash2009,Abadie2010,Baiotti2017}. The neutron-rich ejecta released from BNS/BH-NS mergers undergo a rapid neutron capture nucleosynthesis process ($r$-process) to power kilonovae \citep[e.g.,][]{Li1998,Kasen2013,Just2015,Metzger2019,Nakar2020,Cowan2021}. During the postmerger dynamical time, some of the materials on the accretion disk are accreted into the central compact object and thus produce relativistic jets, which would be detected as GRBs if they point toward the Earth \citep[e.g.,][]{Paczynski1986,Eichler1989,Paczynski1991,Narayan1992,Popham1999}. Then, the interaction between the jet and the external medium might generate long-lasting X-ray, optical, and radio afterglows \citep[e.g.,][]{Rees1992,Berger2005,Nakar2007,Berger2014}. In addition to GWs and electromagnetic radiation, a tremendous number of neutrinos with $\gtrsim 10~\rm MeV$ are also expected to be emitted from the remnant of BNS/BH-NS mergers \citep[e.g.,][]{Sekiguchi2011,Kyutoku&Kashiyama2018}.

GW 170817, originating from a BNS merger, was detected by aLIGO and Virgo on August 17, 2017 \citep{Abbott2017a}. At $\sim 1.7~\rm s$ after the end of the inspiral, GRB 170817A was triggered independently by \emph{Fermi} and \emph{INTEGRAL} \citep[e.g.,][]{Abbott2017b,Goldstein2017,Savchenko2017}. Later, a kilonova AT 2017gfo was also discovered at $\sim 10~\rm h$ after this merger \citep[e.g.,][]{Abbott2017b,Andreoni2017,Arcavi2017,Coulter2017}. More than two different radioactive-powered components were invoked to interpret the observed AT 2017gfo light curves \citep[e.g.,][]{Cowperthwaite2017,Perego2017,Tanaka2017,Villar2017,Tanaka2018,Kawaguchi2018,Wu2019}. An afterglow source coincident with this kilonova from X-ray to radio was subsequently observed \citep[e.g.,][]{Margutti2017,Troja2017,Dobie2018,Lyman2018,Ghirlanda2019}. Neutrino observations with ANTARES, IceCube, the Pierre Auger Observatory \citep{Albert2017}, and Super-Kamiokande \citep{Abe2018} on GW 170817 was conducted. However, no neutrinos were identified. This BNS merger and the subsequent electromagnetic counterparts further constrain the equation of the state (EoS) of NSs \citep[e.g.,][]{Li2017,Li2020}.

Recently, two additional events, GW 190425 \citep{Abbott2020a} and GW 190814 \citep{Abbott2020b}, were found to be associated with compact object mergers; at least one of the objects before mergers is in the lower mass gap, which is strange from what we currently understand as compact binary system formation channels \citep[e.g.,][]{Safarzadeh2020,Zevin2020,Liu2021,Mandel2021}. It is also exciting to note that two BH-NS merger events, i.e., GWs 200105 and 200115, were detected \citep[][]{Abbott2021}. Unfortunately, neither the electromagnetic counterpart nor neutrinos were detected in these events \citep[e.g.,][]{Dobie2019,Ackley2020,Andreoni2020,Coughlin2020,Vieira2020,Watson2020,Zhu2021}.

A stellar-mass BH surrounded by a hyperaccretion disk is one of the widely accepted central engines of BNS/BH-NS mergers or massive collapsars \citep[e.g,][]{Liu2017a, Zhang2018}. Once the disk has a very high accretion rate, the high density and temperature of the disk cause photons to be completely trapped. Only a large number of neutrinos can escape from the disk, releasing the gravitational energy of the BH. Annihilations of a small fraction of neutrinos can launch ultrarelativistic jets. This type of accretion disk is named a neutrino-dominated accretion flow \citep[NDAF, e.g.,][]{Popham1999,Narayan2001,DiMatteo2002,Kohri2002,Kohri2005,Lee2005,Gu2006,Chen2007,Kawanaka2007,Liu2007,Kawanaka2013,Xue2013}; for reviews, see \citet{Liu2017a} and \citet{Zhang2018}. Neutrino emission and GWs induced by the jet precession or anisotropic neutrinos in the NDAF model have been investigated \citep[e.g.,][]{Suwa2009,Romero2010,Sun2012,Liu2016,Liu2017b,Song2018,Song2020,Wei2019,Wei2020}. Besides that, the Blandford-Znajek (BZ) mechanism \citep{Blandford1977} is another way to release energy in BH hyperaccretion scenario.

The anisotropic multimessenger observation of GRBs, kilonovae, neutrinos, and GWs from NDAFs could reveal the BNS/BH-NS merger mechanism, the properties of the outflows, the dynamics of the disk, and the synthesis of the elements. The outline of this paper is as follows. In Section 2, we briefly study NDAFs with outflows. In Section 3, we investigate disk-outflow-driven (DOD) kilonovae, taking into account the anisotropic mass and velocity distributions of the outflows. In Section 4, we present the anisotropic NDAF electron antineutrino spectra by considering the general relativistic effects. In Section 5, we address isotropic GWs and their detectability. A brief summary is provided in Section 6.

\section{NDAFs with outflows}

Disk outflows play an important role in critical accretion systems. We first introduce a well-known relation between the local accretion rate for any radius $\dot{M}$ and the mass supply rates at the outer boundary of the disk $\dot{M}_{\rm outer}$, which can be expressed by a power law form \citep[e.g.,][]{Blandford1999,Yuan2012a,Yuan2014,Sun2019,Liu2020}, i.e.,
\begin{equation}
\dot{M} =\dot{M}_{\rm outer} \left(\frac{R}{R_{\rm outer}} \right)^{p},
\end{equation}
where $p$ is the index parameter that determines the strength of the outflows and $R_{\rm outer}$ is the outer boundary of the disk.

The above relation was first presented by \citet{Blandford1999} for advection-dominated accretion flows (ADAFs). We consider that this power-law form of the local accretion rate can also be applied to the NDAF models. The reasons are as follows. First, \citet{Liu2012a} investigated the vertical distribution of the Bernoulli parameter in NDAFs, we found that in the outer region the Bernoulli parameter is positive, therefore, outflows might be formed in this region. Actually, many numerical simulations on NDAFs showed that the disk outflows are indispensable \citep[e.g.,][]{Fernandez2013,Fernandez2015,Just2015,Wu2016,Lippuner2017,Siegel2017}. Second, NDAFs are the naturally extended branch of slim disks for very high accretion rates in the well-known $\dot{M}-\Sigma$ plane \citep{Liu2017a}, where $\Sigma$ and $\dot{M}$ are the surface density of the disk and the accretion rate, respectively. Meanwhile, the outer region of NDAFs is indeed advection-dominated \citep[e.g.,][]{Popham1999,Liu2007}. Third, the efficiency of neutrino emission in the systems of Schwarzschild BHs surrounded by NDAFs without outflows is generally estimated, i.e., $L_{\nu } /\dot{M} c^{2} \lesssim 0.035$ for $\dot{M} \lesssim 1~M_\odot~\rm s^{-1}$ \citep[e.g.,][]{Kawanaka2007}, where $L_{\nu}$ is the total neutrino luminosity. For the case of NDAF with outflows, the values of the above efficiency are in the range between a few percent and nearly 10$\%$ for all cases of \citet{Just2015}. This result indicates that the NDAF with outflows is radiatively inefficient, which is similar to the ADAF model. The NDAF outflows are widely applied to the nucleosynthesis and dynamic motions in collapsar and merger scenarios \citep[e.g.,][]{Surman2008,Surman2011,Song2018,Song2019,Liu2019,Liu2020}. Outflows are generated in the advection-dominated region, which leads to the low net accretion rate in the inner region, thus the neutrino-cooling-dominated region should be very narrow. Hence we roughly adopted the above power-law form to describe the outflows of the whole NDAFs.

In our calculation, the inner boundary and the outer boundary of the disk are taken as $R_{\rm inner}=2.3~R_{\rm g}$ (for the BH mass $M_{\rm BH}=3~M_\odot$) and $R_{\rm outer}=50~R_{\rm g}$, respectively, where $R_{\rm g}=G M_{\rm BH}/c^{2}$ is the Schwarzschild radius. The detailed dynamic equations are adopted from \citet{Liu2010} and \citet{Sun2012}, from which the structure of a steady and axisymmetric NDAF can be calculated.

\begin{figure}
\centering
\includegraphics[width=0.95\linewidth]{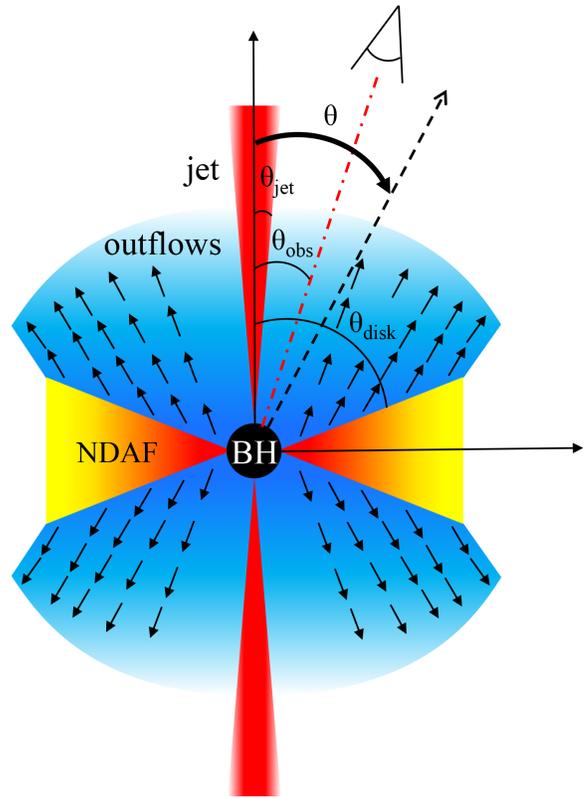}
\caption{Schematic diagram of NDAFs with outflows.}
\label{fig:kilonovae cartoon}
\end{figure}

The total pressure $P$ consists of four terms, namely, the gas pressure, the radiation pressure, the electron degeneracy pressure, and the neutrino pressure \citep[e.g.,][]{Kohri2005,Liu2007}, which can be written as
\begin{equation}
P=P_{\rm gas} + P_{\rm rad} + P_{\rm e} + P_\nu,
\end{equation}
and the energy balance equation is expressed as
\begin{equation}
Q_{\rm vis}^{+}= Q_{\rm adv}^{-}+ Q_{\rm photondis}^{-} + Q_{ \rm \nu }^{-},
\end{equation}
where $Q_{\rm vis}^{+}$ is the viscous heating rate and $Q_{\rm adv}^{-}$, $Q_{\rm photondis}^{-}$, and $Q_{\rm \nu}^{-}$ are the advective cooling rate, the photodisintegration of the $\alpha$-particle cooling rate, and the neutrino loss cooling rate, respectively \citep[e.g.,][]{Kohri2005,Liu2007,Xue2013}. The term $Q_{\rm photondis}^{-}$ is always ignored because it only show the little effect in a narrow middle region of the disk \citep[e.g.,][]{Janiuk2004,Liu2007}. The details of the neutrino physics related to the above equations are adopted from \citet{Liu2017a}. It should be mentioned that the local net accretion rate $\dot{M}$ is applied, so there are no terms of the local outflow pressure and cooling rate in the EoS and the energy equation. Thus, they only depend on the difference in the total pressure and viscous heating rates between two adjacent radii \citep{Liu2020}. This method is appropriate in the numerical calculations, which has been further verified by comparing the Bernoulli parameters \citep{Liu2012a} and our primary NDAF simulations.

\citet{Liu2015} showed that in most short-duration GRB (SGRB) cases, the fall back remnant mass is less than $\sim 0.5~M_{\odot}$ and can reach $\sim 0.8~M_{\odot}$ in BH-NS merger cases \citep{Liu2012b}. The total remnant mass of BNS mergers is generally smaller than that of BH-NS mergers, which depends on the EoS of NSs, the mass ratio of the binary, the total binary mass, and the orbital period \citep[e.g.,][]{Oechslin&Janka2006,Foucart2012,Foucart2014,Dietrich2015,Just2015,Kiuchi2015}. In the following,we give four typical cases, i.e., ($p, ~(M_{\rm disk} + M_{\rm outflow}) / M_{\odot}$) = (0.3, 0.2), (0.8, 0.2), (0.3, 0.5), and (0.8, 0.5), where $M_{\rm disk}$ is the time-independent disk mass and $M_{\rm outflow}$ is the total net outflow mass (see Table 1). The sum of these masses represents the total fall back remnant mass around the BH. The disk mass can be constrained by $M_{\rm disk} =4 \pi\int_{R_{\rm inner} }^{R_{\rm outer}} \rho R H dR$, where $\rho$ and $H$ are the density and half-thickness of the disk, respectively. Note that $p$ = 0.3 and 0.8 represent the weak and strong disk outflows, and $M_{\rm disk} + M_{\rm outflow}$ = 0.2 and $0.5~M_{\odot}$ denote the fall back remnant mass in the cases of BNS and BH-NS mergers, respectively. Here, we set the dimensionless BH spin parameter $a_{*}=0.9$ since the post-merger BH is generally a fast rotator \citep[e.g.,][]{Popham1999,Janiuk2008,Liu2015,Song2015}, and the viscous parameter of the disk $\alpha=0.1$.

\section{DOD kilonovae}
\subsection{Methods}

The neutron-rich material ejected during compact object mergers can produce elements heavier than iron through the $r$-process \citep[e.g.,][]{Lattimer1974,Symbalisty1982}. \citet[]{Li1998} first noted that the radioactive decay of these elements could feed the supernova (SN)-like thermal transient source. \citet[]{Metzger2010} calculated the luminosity of an optical-infrared thermal transient source from a BNS merger and named it ``kilonova''. \citet{Tanaka2013} and \citet{Barnes2013} showed that lanthanides have a higher opacity than iron. In addition, there are several main additional energy sources, such as the decay of free neutrons \citep[e.g.,][]{Metzger2015} and the shock heating from the collision of relativistic jets with the ejecta \citep[e.g.,][]{Bucciantini2012,Gottlieb2018,Piro2018}. Unlike the roughly isotropic distribution of the ejecta from BNS mergers, the dynamical ejecta from BH-NS mergers predominantly distribute over the orbital plane in the shape of a crescent, with a small angle in the vertical direction and sweeping across half the plane in the horizontal disk direction \citep[e.g.,][]{Kyutoku2013,Kyutoku2015,Kawaguchi2016,Zhu2020}. In addition to the ejecta from mergers, winds from a long-lived magnetar \citep[e.g.,][]{Yu2013,Yu2018,MetzgerPiro2014,Ren2019} or the outflows from an accretion disk \citep[e.g.,][]{Just2015,Wu2016,Perego2017,Siegel2017,Song2018} can be considered the other significant ejecta source to power the luminous kilonovae. During the past few years, several SGRBs associated with kilonova candidates were searched, i.e., GRBs 130603B \citep[e.g.,][]{Berger2013,Tanvir2013}, 050709 \citep[e.g.,][]{Jin2016}, 060614 \citep[e.g.,][]{Jin2015,Yang2015}, and 070809 \citep[e.g.,][]{Jin2020}.

In this section, we revisit the DOD kilonova light curves in different optical bands. NDAF outflows with anisotropic mass and velocity distributions might produce anisotropic DOD kilonovae. Due to the geometry of outflows, we sum the individual contributions of materials in a solid angle $\Delta \Omega$ to compute light curves. Figure 1 shows a schematic diagram of a BH surrounded by an NDAF with outflows.

\citet{Yuan2012b,Yuan2015} simulated the mass and velocity distributions of the outflows from the hot accretion flows with Eddington accretion rate. The difference on the accretion rates between NDAFs and hot accretion disks is about 16 orders of magnitude. Since NDAFs are similar to the radiatively inefficient ADAFs on the dynamics \citep[e.g.,][]{Just2015}, we adopt their results to build the outflow distributions from NDAFs, which are also briefly verified by our primary simulations. The mass distribution of the outflow material can be defined as
\begin{equation}
M(\theta)=M_{0}~(\frac{\theta}{\theta_0})^{k}, ~\theta_{\rm jet}< \theta < \theta_{\rm disk},
\end{equation}
where $\theta_{\rm jet}$ and $\theta_{\rm disk}$ are respectively the half-opening angles of the jet and the disk in the first quadrant as shown in Figure 1, and then we set $\theta_0=\theta_{\rm jet}$. The index $k \approx$ 3.72, which is fitted from the two-dimensional simulations in \citet{Yuan2012b,Yuan2015}. Here we assume that the outflows escape from the disk independent of the radius, because their distribution is not sensitive for the far enough cases. The mass normalization is determined by
\begin{equation}
M_{\rm outflow} = 2\int_{\theta_{\rm jet}}^{\theta_{\rm disk}}\int_{0}^{2\pi}M(\theta)\rm sin\theta d\theta d\varphi,
\end{equation}
from which can obtain the value of $M_0$. The outflows are composed of blobs with the same velocities, so $v(\theta)$ is written as
\begin{equation}
v(\theta)=v_{0}~(\frac{\theta}{\theta_0})^{s}, ~\theta_{\rm jet}< \theta < \theta_{\rm disk}
\end{equation}
with $s \approx -1.13$ and $v_0 \approx 0.47~c$. Obviously, the rangeability of the outflow velocity is much smaller than that of the outflow mass. It should be mentioned that the range of $\theta$ should exclude the jet and disk opening angles as shown in Figure 1.

The material of outflows at a certain solid angle $M(\theta) \Delta \Omega$ is divided into $N-1$ layers with a constant expansion velocity $v(\theta)$. Here, we define the mass layers from the bottom $R_{\rm min} (\theta)=v (\theta) t$ to the head layers $R_{\rm max} (\theta) = R_{0} + v (\theta) t$ by the subscript $i=1,2,...,N-1$, where $R_{0}=10^8~\rm cm$ (much larger than the outer boundary of the disk) is the initial radius conditions of the outflows. The location of the $i$th layer in a solid angle at time $t$ is $R_{i} (\theta)$ as shown in Figure 2.

\begin{figure}
\centering
\includegraphics[width=0.95\linewidth]{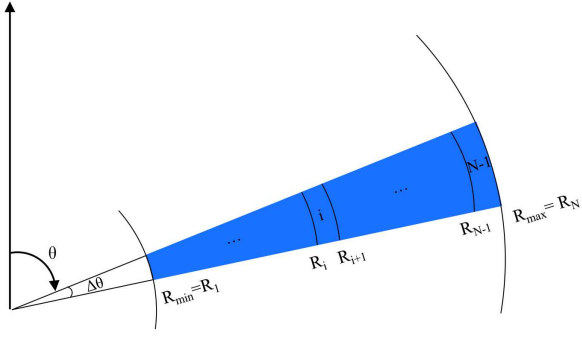}
\caption{Schematic diagram of ejecta divided into $N-1$ layers.}
\end{figure}

For a matter with the same velocity at a solid angle $\Delta \Omega$, the mass of the $i$th layer is
\begin{equation}
m_{i}= M \frac{R^{3} _{i+1} - R^{3} _{i} }{R^{3} _{\rm max}-R^{3} _{\rm min} } \Delta \Omega.
\end{equation}

By improving a unified kilonova model \citep[e.g.,][]{Metzger2019}, the thermal energy $E_{i}$ evolution of the $i$th layer at $\Delta \Omega$ is expressed as
\begin{equation}
\frac{dE_{i}} {dt}= m_{i} \dot{q}_{\rm r} \eta_{{\rm th}} - \frac{ E_{i}}{R_{i}}{dR_{i}\over dt}-L_{i},
\end{equation}
where $\dot{q}_{\rm r}$ is the radioactive energy supplied per unit mass, $\eta_{\rm th}$ is the thermal efficiency of the radioactive energy supply, and $L_i$ is the observed luminosity contributed by this layer. The three terms on the r.h.s in order are the energy produced by the radioactive heavy element decay, the adiabatic expansion cooling, and the radiation cooling in each layer.

The radioactive energy supply per unit mass can be written as \citep[e.g.,][]{Korobkin2012}
\begin{equation}
\dot{q}_{\rm r}=4 \times 10^{18}\left[{1\over2}-{1\over \pi}{\rm
arctan}\left({{t-t_0}\over \sigma}\right)\right]^{1.3}\rm
erg~s^{-1}g^{-1},
\end{equation}
where $t_0=1.3$~s and $\sigma=0.11$~s. Its thermalization efficiency is expressed as \citep[e.g.,][]{Barnes2016}
\begin{equation}
\eta_{\rm th}=0.36\left[\exp(-0.56 t_{\rm day})+{\ln (1+0.34t_{\rm day}^{0.74})\over 0.34t_{\rm day}^{0.74}}\right],
\end{equation}
where $t_{\rm day}=t/1\rm day$.

The observed luminosity released of the $i$th layer at $\Delta \Omega$ is
\begin{equation}
L_{i}={E_{i}\over \max[t_{{\rm d},i},t_{{\rm lc},i}]},
\end{equation}
where $t_{{\rm d},i}$ is the radiation diffusion timescale of the $i$th layer, and $t_{{\rm lc},i}=R_{i}/c$ represents the light-crossing time.

The radiation diffusion timescale is described as
\begin{equation}
t_{{\rm d},i}=\frac{3\kappa}{\Delta \Omega R_{i} c}{\sum\limits_{j=i}^{N-1}}m_{j}.
\end{equation}

Finally, by summing the radiation luminosity contributed of all layers at $\Delta \Omega$, the total bolometric luminosity $L_{\rm bol}$ of outflows at $\Delta \Omega$ is given as
\begin{equation}
L_{\rm bol}={\sum\limits_{i=1}^{N-1}} L_{i}.
\end{equation}.

\begin{figure*}
\centering
\includegraphics[width=0.95\linewidth]{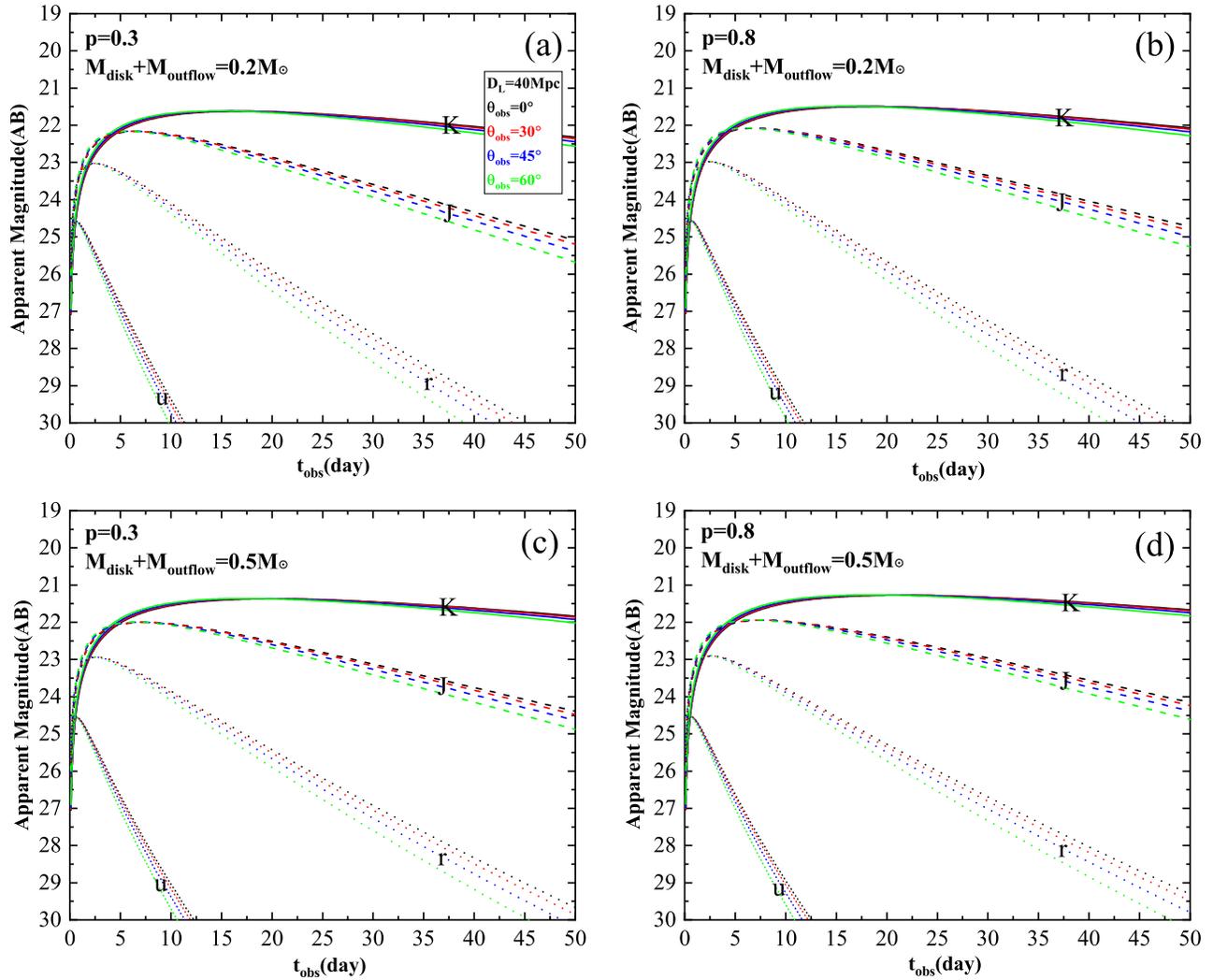}
\caption{Viewing-angle-dependent urJK-band light curves of DOD kilonovae at a distance of 40 Mpc for the cases of ($p, ~(M_{\rm disk} + M_{\rm outflow}) / M_{\odot}$) = (0.3, 0.2), (0.8, 0.2), (0.3, 0.5), and (0.8, 0.5). The black, red, blue, and green lines denote the viewing angles $\theta_{\rm obs}=$ $0^{\circ}$, $30^{\circ}$, $45^{\circ}$, and $60^{\circ}$, respectively. The solid, dashed, thin dotted, and thick dotted lines represent the K, J, r, and u bands, respectively.}
\label{fig:kilonova}
\end{figure*}

Taking into account the blackbody radiation spectrum, the radiation of the ejecta is always assumed to originate from the photosphere $R_{\rm ph}$ and the effective temperature of each point of the DOD kilonova emission can be expressed as
\begin{equation}
T_{\rm eff}=\left({ L_{\rm bol}\over\sigma_{\rm SB}\rm \Delta \Omega \emph{R}_{\rm ph}^2}\right)^{1/4},
\end{equation}
where $\sigma_{\rm SB}$ is the Stephan-Boltzmann constant and the photosphere radius $R_{\rm ph}$ is defined as that of the mass layer beyond which the optical depth equals unity. The ejecta is only distributed between $R_{\rm min}$ and $R_{\rm max}$, so we have to adopt such an approximate method that if the total optical depth of the ejecta $\tau_{\rm tot} \leq 1$, the photosphere radius $R_{\rm ph}$ is the minimum radius $R_{\rm min}$ \citep[e.g.,][]{Yu2018,Ren2019}; if the optical depth of the $(N-1)$th layer $\tau_{N-1} > 1$, $R_{\rm ph}=R_{\rm max}$. The total optical depth is $\tau_{\rm tot}= {\sum\limits_{i=1}^{N-1}} \tau _{i}$, $\tau_{i}$ is the optical depth of the $i$th layer and can be described as $\tau _{i} =\int_{R_{i} }^{R_{i+1} } \kappa \rho_{\rm outflow} dR$, where $\rho_{\rm outflow}$ is the outflow density and can be derived from the outflow distribution. The value of the average effective gray opacity $\kappa \sim 0.5\mbox{\,--\,}30\mathrm{\ cm^{2}\ g^{-1}}$ represents a wide range from lanthanum-free to lanthanum-rich matter \citep[e.g.,][]{Tanaka2020}. Here, $\kappa=20~\rm{cm^{2} ~g^{-1}}$ is adopted.

Subject to the propagation of light, the photons are emitted at a given time $t$ and reach the observer at time
\begin{equation}
t_{\rm obs} =t + \frac{[R_{\rm ph} (\theta_{\rm obs}) - R_{\rm ph}(\theta)]\rm \cos\Delta\theta}{c},
\end{equation}
where $\Delta\theta$ is the angle between the direction of movement of the point on the photosphere layer and the direction of the line of sight.

The distribution of the observed flux with the solid angle is
\begin{equation}
dF_{\nu_{\rm }}(\nu,t_{\rm obs})={2\pi h\nu^3\over c^2}{1\over1-\exp(h\nu/k_{\rm B}T_{\rm eff})}{R_{\rm ph}^2 d\Omega \over 4\pi D_{\rm L}^2},
\end{equation}
where $k_{\rm B}$ is the Boltzmann constant, $h$ is the Planck constant, and $D_{\rm L}$ is the luminosity distance.

According to the half-day arc equation, i.e.,
\begin{equation}
\varphi(\theta)=\rm arccos (-\rm cot \theta \rm cot \theta_{\rm obs} ),
\end{equation}
the range of longitude visible to the observer in a certain $\theta$ is $[-\varphi(\theta), \varphi(\theta)]$ if the observer is located at $\varphi = 0 ^{\circ}$.

The observed flux density of the DOD kilonova emission from a certain viewing angle $\theta_{\rm obs}$ in the first quadrant can be written as
\begin{equation}
F_{\nu_{\rm }}(\nu,t_{\rm obs})= 2\int_{\theta _{\rm jet} }^{\theta _{\rm obs} +\frac{\pi }{2} } \int_{0}^{\varphi (\theta) }  dF_{\nu_{\rm }}.
\end{equation}

In our calculation, we set $\theta_{\rm jet}=5 ^{\circ}$, which corresponds to the typical value of the half-opening angles of jets; $\theta_{\rm disk}=80 ^{\circ}$, which is adopted from the typical half-opening angle of the inner region of NDAFs \citep[e.g.,][]{Popham1999,Liu2007}, and taken into account that the outflows far away from the disk might diffuse in the $\theta$-direction; and $D_{\rm L}$ = 40 Mpc. Finally, we can express the monochromatic AB magnitude $M_{\nu}$ by the relation of $M_{\nu}=-2.5\rm log_{10}(\emph{F}_{\nu}/3631\rm Jy)$.

\subsection{Results}

Figure~\ref{fig:kilonova} shows the viewing-angle-dependent urJK-band light curves of DOD kilonovae at a distance of $40~\rm Mpc$. The black, red, blue, and green lines denote the viewing angles $\theta_{\rm obs}=$ $0^{\circ}$, $30^{\circ}$, $45^{\circ}$, and $60^{\circ}$, respectively. The solid, dashed, thin dotted, and thick dotted lines represent the K, J, r, and u bands, respectively. One can find that the DOD kilonova luminosity decreases as the viewing angle $\theta_{\rm obs}$ increases. These properties appear in the J, u, and r bands but in the K band. This is because the flux at different solid angles is related to the position of the photosphere radius. The minimum (or maximum) values of the outflow mass (or velocity) appear near the pole direction. Thus the results show that the anisotropy is not very sensitive to the mass distribution and mainly determined by the velocity distributions.

As comparisons among four figures, the values of $p$ and $M_{\rm disk}+M_{\rm outflow}$ have little effects on the luminosities for a certain viewing angle. In other words, there is slight difference on the light curves in both the BNS and BH-NS merger cases, which is caused by the value of the outflow velocity rather than that of the outflow mass. The typical magnitudes are dimmer than $\sim 24$ mag in the ultraviolet band and dimmer than $\sim 21~\rm mag$ in the infrared band. The emission peaks appear at $\sim 1$ day in the u-band and $\sim 1$ week in the K-band. The DOD kilonova emission is dominated by the K-band after $\sim 5$ days. The mean power injected by the $r$-process in the total outflows can be obtained by $\bar{P}_{r} = \int_{0}^{50 ~\rm days} M_{\rm outflow} \dot{q}_{\rm r} \eta_{{\rm th}} dt / (50~ \rm days)$, and the mean bolometric luminosity of the total outflows is calculated by $\bar{L}_{\rm bol}=\int_{0}^{50 ~\rm days}\int_{0}^{4\pi}L_{\rm bol}d\Omega dt/(50~ \rm days)$, which are shown in Table 1. We find that the typical DOD kilonovae are dimmer than AT 2017gfo associated with BNS mergers. Since the distributions of the outflow mass and velocity are inversely correlated with $\theta$, it is difficult for the DOD kilonovae to be dominant in AT 2017gfo. This is consistent with previous kilonova models where AT 2017gfo is composed of different components and mainly powered by the dynamical ejecta \citep[e.g.,][]{Cowperthwaite2017,Kasen2017,Villar2017}. For the case of BH-NS mergers, the dynamical ejecta are mainly on the orbital plane, so the role of NDAF outflows should be more pronounced and they might become the dominant components. In other words, the DOD kilonovae might be unobservable in the BNS mergers, but can be detected in the BH-NS mergers occurred in the nearby universe. If a DOD kilonova is associated with GW 200105 at a distance of $\sim 300$ Mpc, their K-band peak luminosities at AB magnitudes are less than 26 mag; thus, they are too faint to be detected by the telescopes performing the observation tasks on GW electromagnetic counterparts. Moreover, the DOD kilonova has at least a natural blue and red component due to the anisotropic outflow mass and velocity distribution from the pole to equator directions.

Unlike the roughly isotropic distribution of the ejecta from BNS mergers, the dynamical ejecta from BH-NS mergers predominantly distribute over the orbital plane in the shape of a crescent, with a small angle in the vertical direction and sweeping across half the plane in the horizontal disk direction

Neutrino irradiation may increase the electron fraction $Y_{e}$ of neutron-rich material \citep[particularly $\nu_{e} + n \rightarrow p + e^{-}$, e.g.,][]{Korobkin2012,Wanajo2014,Kyutoku&Kashiyama2018}. We reasonably ignore this effect because the outflows are mainly launched from the outer region of the disk, where only a small number of neutrinos are produced, and the neutrino optical depth of the disk outflows is generally thin. From the distributions of the outflow density and temperature, we can estimate the sum of absorptive and scattering optical depths of neutrinos \citep[e.g.,][]{DiMatteo2002,Gu2006} escaped along the vertical direction at $30~R_{g}$, $\sim 2 \times 10^{-8}$.

\section{MeV neutrinos}
\subsection{Methods}

\begin{figure*}
\centering
\includegraphics[width=0.95\linewidth]{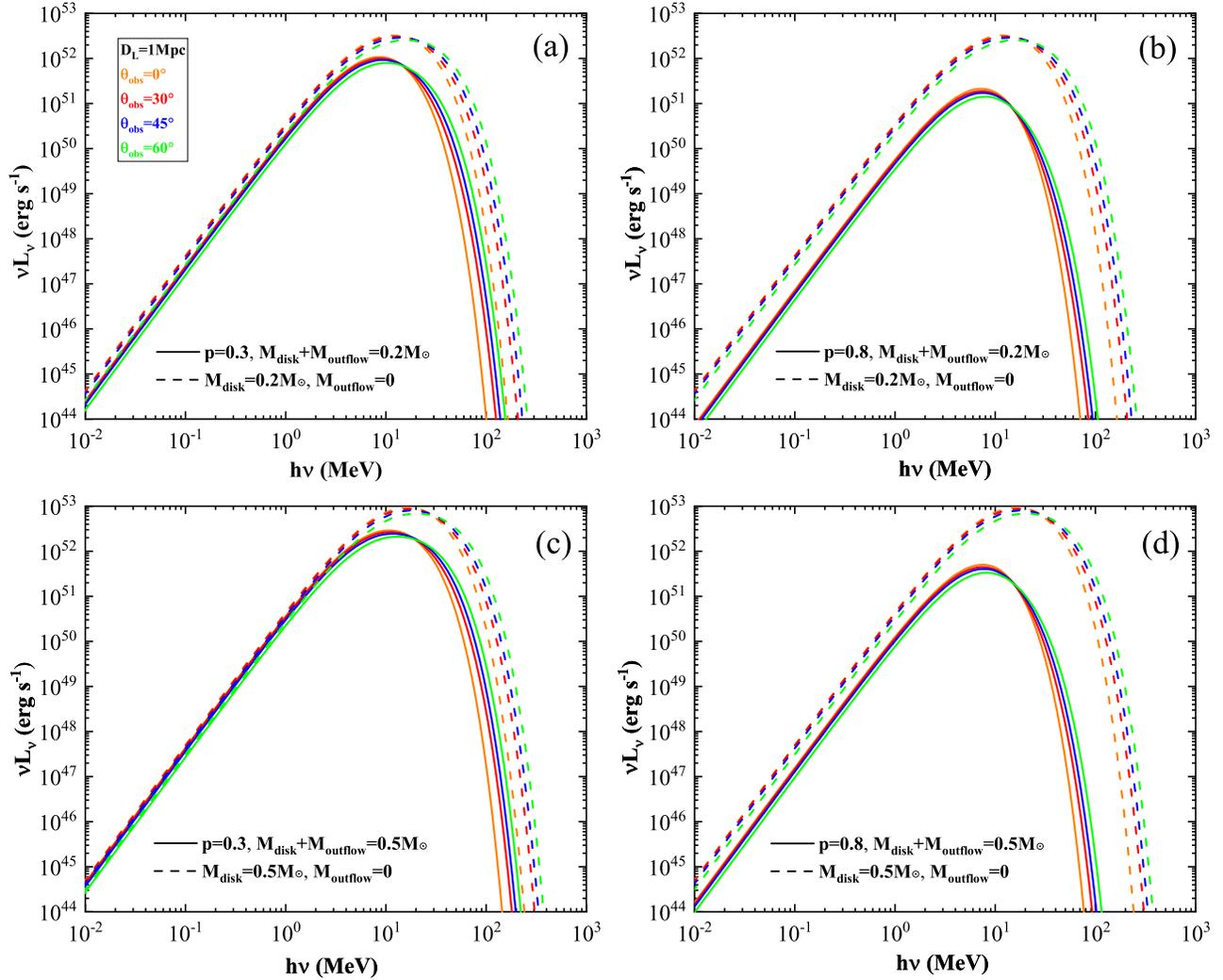}
\caption{Viewing-angle-dependent electron antineutrino spectra of NDAFs with (solid lines) or without (dashed lines) outflows in merger scenarios at a distance of $1~\rm Mpc$. The orange, red, blue, and green lines correspond to $\theta_{\rm obs}$ = $0^{\circ}$, $30^{\circ}$, $45^{\circ}$, and $60^{\circ}$, respectively.}
\label{fig:neutrino}
\end{figure*}

MeV neutrinos from core-collapse SNe \citep[CCSNe, e.g.,][]{Friedland2010,Warren2020} and their central NDAFs \citep[e.g.,][]{Liu2016,Wei2019,Song2020} have been investigated. Although the neutrino luminosity of NDAFs is $1-3$ orders of magnitude lower than that of CCSNe, a certain number of neutrinos could still be detected in the Local Group by prospective detectors \citep[e.g.,][]{Liu2016}. In this work, the anisotropic neutrino emission from NDAFs in binary compact object mergers is calculated.

We employ ray-tracing methods \citep[e.g.,][]{Fanton1997,Li2005} to account for the propagation of neutrinos and calculate the observed neutrino energy spectrum, which is primarily influenced by the general relativity under the strong gravity of the central BH. Here, we assume that the equatorial plane is the location of the main source of neutrinos, the neutrinos are emitted isotropically from all radii of the disk, and the disk thickness is neglected.

The neutrino trajectories from the emitter must satisfy the geodesic equation \citep{Carter1968}. Therefore, the location of the launch point of the disk can be tracked by each pixel point of the observed image. The neutrino energy redshift can be calculated by integrating the velocity and gravitational potential at the corresponding emission position point. The observed neutrino flux is obtained by integration of all pixel points \citep[e.g.,][]{Liu2016,Wei2019}, i.e.,
\begin{equation}
F_{{E}_{\rm{obs}}}=\int_{\rm image}g^{3}I_{E_{\rm{em}}}d\Omega _{\rm{obs}},
\end{equation}
where $g \equiv E_{\rm obs} / E_{\rm em}$ is the energy shift factor, $E_{\rm{em}}$ is the neutrino emission energy from the local disk, $E_{\rm{obs}}$ is the observed neutrino energy, and $\Omega _{\rm{obs}}$ is the solid angle of the disk image with respect to the observer. Based on the cooling rate via electron antineutrinos $Q_{\bar{\nu}_e}$, the local emissivity $I_{E_{\rm{em}}}$ is obtained by
\begin{equation}
I_{E_{\rm{em}}}=Q_{\bar{\nu}_e}\frac{F_{E_{\rm{em}}}}{\int F_{E_{\rm{em}}}dE_{\rm{em}}},
\end{equation}
where $F_{E_{\rm{em}}}=E_{\rm{em}}^{2}/[\exp (E_{\rm{em}}/kT-\eta )+1]$ is the unnormalized Fermi-Dirac spectrum \citep[e.g.,][]{Rauch1994,Fanton1997,Li2005,Liu2016}. Therefore, the luminosity distribution can be derived as
\begin{equation}
L_{\rm{\nu} }=4\pi D_{\rm L}^{2} F_{{E}_{\rm{obs}}}.
\end{equation}

It should be mentioned that electron antineutrinos can be detected via the inverse beta decay (IBD) reaction, and the IBD is the best channel for the most of present and upcoming neutrino detectors. Thus we discuss the electron antineutrino spectra here.

\subsection{Results}

Figure~\ref{fig:neutrino} shows the effects of the viewing angles on the electron antineutrino spectra of NDAFs with different outflow strengths (solid lines) and without outflows (dashed lines) in the BNS/BH-NS merger scenarios at a distance of $1~\rm Mpc$. The orange, red, blue, and green lines correspond to the viewing angles $\theta_{\rm obs}$ = $0^{\circ}$, $30^{\circ}$, $45^{\circ}$, and $60^{\circ}$, respectively. In the case of NDAFs with different outflow strengths, the neutrino energies are generally in the range of $\sim 10~\rm keV-100 ~\rm MeV$, and the peaks are at approximately $10-20$ MeV. The corresponding neutrino luminosities of NDAFs reach $\sim 2\times10^{51}-3\times10^{52}~\rm erg~s^{-1}$. However, in the case of NDAFs without outflows, the peaks of neutrino energies move slightly to the higher values, and the maximum neutrino luminosity can be $\sim 3\times10^{52}-9\times10^{52}~\rm erg~s^{-1}$. Moreover, we find that the high-energy ($\gtrsim 10~\rm MeV$) neutrinos have a more anisotropic luminosity, but the low-energy ($\lesssim 10~\rm MeV$) neutrinos are approximately isotropic. High-energy neutrinos are mainly produced in the inner region of the disk, which means that they are significantly deflected to a large viewing angle by BH gravity. Therefore, the high-energy neutrino luminosities increase with increasing viewing angles. In contrast, low-energy neutrinos are mainly generated in the outer region of the disk, so the general relativity has weak effects on them. Comparing the weak and strong outflows of NDAFs in BNS or BH-NS merger cases, we find that the neutrino luminosity decreases with increasing outflow strength. Actually, there is a natural competition between the masses and energies of the outflows and inflows \citep[e.g.,][]{Song2018,Liu2020}.

The effects of neutrino oscillations are not taken into account, which not only corrects the neutrino energy specta but also decreases the detectable neutrino fluxes by a factor of up to $\sim 2-3$ \citep[e.g.,][]{Friedland2010,Duan2011,Cherry2012,Liu2016}. \citet{Kyutoku&Kashiyama2018} proposed that monitoring $\gtrsim2500$ merger events within $\lesssim200$ Mpc identified by GW detectors could give us a chance of detecting a single neutrino for the merger rate of $1~\rm Mpc^{-3}~\rm Myr^{-1}$ with a human time-scale operation of $\sim80$ Myr. MeV neutrinos have not been detected in any of the recent binary compact object mergers, probably because they are too far away to be detected. The detectable neutrinos from NDAFs in merger events also need strict requirements on the distance (or the brightness), which is at least similar to the possibly detectable distance of NDAFs in the center of CCSNe, $\lesssim$ 1 Mpc, i.e., location in the Local Group \citep[see e.g.,][]{Liu2016}. Moreover, as mentioned in the above section, we ignore the interaction between neutrinos escaped from the disk and the high-velocity disk outflows.

\begin{figure}
\centering
\includegraphics[width=1.1\linewidth]{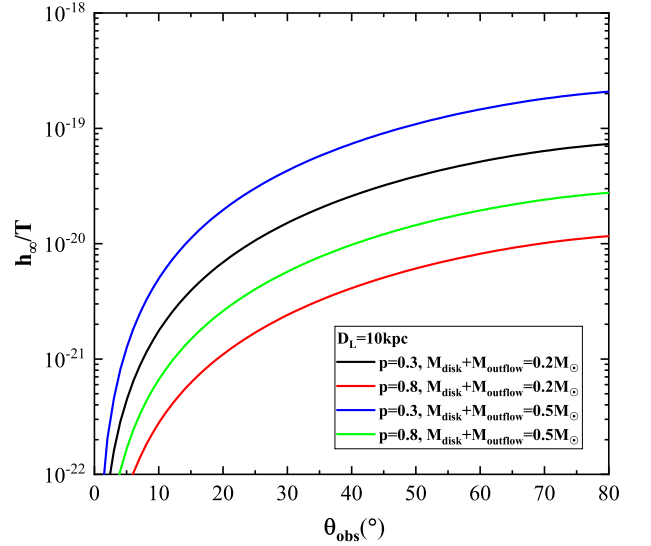}
\caption{$h_{\infty}/T$ as a function of the viewing angle $\theta_{\rm obs}$ at a distance of $10$ kpc for four cases.}
\label{fig:GW1}
\end{figure}

\section{GWs}
\subsection{Methods}

\citet{Epstein1978} first proposed GW radiation resulting from anisotropic neutrinos, and the methods have been applied to CCSNe \citep[e.g.,][]{Burrows1996,Kotake2006,Kotake2007}. GWs arising from anisotropic neutrino emission in NDAFs have been investigated \citep[e.g.,][]{Suwa2009,Kotake2012,Liu2017b,Song2020,Wei2020}. \citet{Liu2017b} examined the BH spin and accretion rate impacts on the strains of GWs in NDAFs; compared their strains, frequencies, and detection possibility to those from magnetars; and presented no GW emission from the BZ mechanism \citep{Blandford1977}. The above works mainly focus on the collapsar scenario, here we adopt the similar method to calculate the GWs from NDAFs in BNS/BH-NS merger scenarios.

For long-lived neutrino emissions, the GW amplitude converges to a nonvanishing value $h_{\infty}$. It is related to the viewing angle $\theta_{\rm obs}$ and is derived as follows \citep[e.g.,][]{Suwa2009}:
\begin{eqnarray}
h_{\infty} (\theta_{\rm obs})=\frac{2G (1+2\cos \theta_{\rm obs})}{3 c^4 D_{\rm L}}\tan^2\left(\frac{\theta_{\rm obs}}{2}\right) \bar{L}_{\nu}T, \nonumber\\
\end{eqnarray}
where $\bar{L}_{\nu}=2 \pi \int_0^T \int_{R_{\rm inner}}^{R_{\rm outer}} Q^-_\nu RdR dt /T$ and $T$ are the mean neutrino luminosity above (or below) the disk (see Table 1) and the center engine activity duration, respectively.

Considering an SGRB powered by an NDAF as one single pulse, we construct the luminosity of neutrino evolution of time $L_{\nu}(t)$ as \citep[e.g.,][]{Suwa2009}
\begin{eqnarray}
L_{\nu} (t)= \bar{L}_{\nu} \Theta (t) \Theta (T-t),
\end{eqnarray}
where $\Theta$ is the Heaviside step function. Central engine intermittent activity should also be considered. This motivates us to list the case of multiple pulses,
\begin{eqnarray}
L_{\nu}(t)=\sum_{i=1}^{N'} \frac{\bar{L}_{\nu}T}{N'\delta t} \Theta (t-\frac{i}{N'} T) \Theta (\frac{i}{N'} T+\delta t-t),
\end{eqnarray}
where $N'$ is the number of pulses and $\delta t$ is the duration of one pulse. $N'\delta t$ should be shorter than $T$, except in the case of a single pulse of $N'=1$.

\begin{figure*}
\centering
\includegraphics[width=0.95\linewidth]{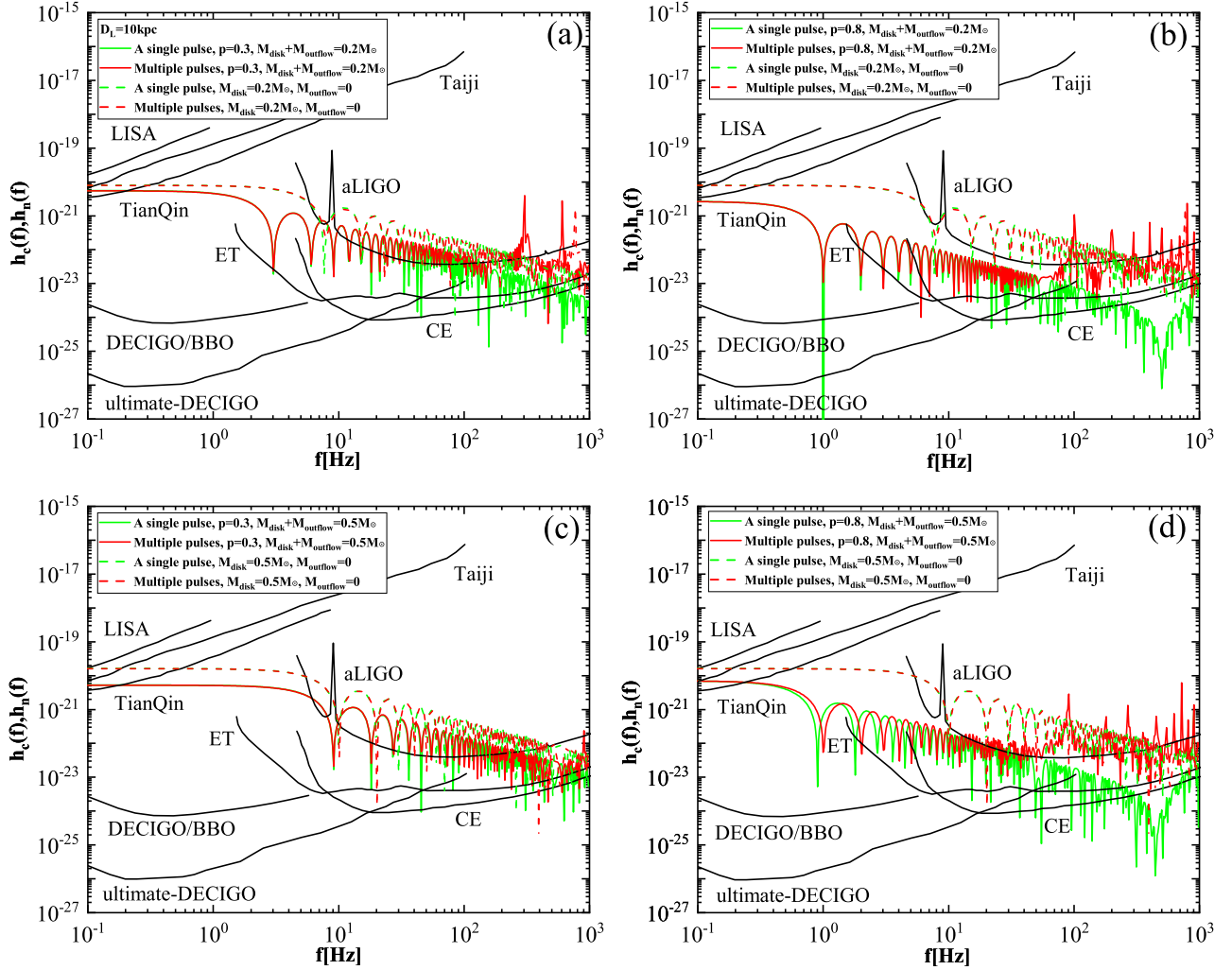}
\caption{Characteristic strains of GWs from NDAFs with (solid lines) or without (dashed lines) outflows in BNS/BH-NS merger scenarios at a distance of $10$ kpc. The green and red lines represent the cases of SGRBs with a single pulse and the multiple pulses, respectively. The black solid lines represent the sensitivity lines (the noise amplitudes $h_{\rm n}$) of aLIGO, ET, CE, LISA, Taiji, TianQin, DECIGO/BBO, and ultimate-DECIGO.}
\label{fig:GW2}
\end{figure*}

$L_{\rm{\nu}}(t)$ is expressed as an inverse Fourier transform into the form,
\begin{eqnarray}
L_{\rm{\nu} }(t)=\int_{-\infty }^{+\infty} \tilde{L}_{\rm{\nu} }(f)e^{-2\pi ift}df,
\end{eqnarray}
for a given frequency $f$.

The nonvanishing GW amplitude of NDAFs $h_+(t)$ is given by \citep[see e.g.,][]{Mueller1997}
\begin{eqnarray}
h_+(t)=\frac{2 G}{3 D_{\rm L} c^4} \int^{t-D_{\rm L}/c}_{-\infty} L_\nu (t')d t',
\end{eqnarray}
and the local energy flux of GWs can be estimated by \citep[e.g.,][]{Suwa2009}
\begin{eqnarray}
\frac{dE_{\rm GW}}{D_{\rm L}^2 d\Omega dt}= \frac{c^3}{16 \pi G} | \frac{d}{dt} h_+(t)|^2.
\end{eqnarray}

\begin{table*}
\begin{center}
\caption{Key properties of multimessenger signals.}
\begin{tabular}{clcccc}
\hline
\hline
&Luminosity (or power) &  Case I & Case II & Case III & Case IV \\
\hline
&$M_{\rm outflow} /  M_{\odot}$       & 0.12      & 0.19    &  0.31    &0.44  \\
&$\bar{P}_{r} / 10^{44}~\rm erg~s^{-1}$       & 0.46      & 0.72    &  1.15    &1.67  \\
&$\bar{L}_{\rm bol} / 10^{39}~\rm erg~s^{-1}$    & 4.93      & 5.77    &  6.76    &7.56  \\
&$\bar{L}_{\rm{\nu} } / 10^{53}~\rm erg~s^{-1}$  & 0.43      & 0.07    &  1.23    &0.16  \\
&$\bar{P}_{\rm GW} / 10^{45}~\rm erg~s^{-1} $   & 0.88      & 0.07    &  2.38    &0.43  \\
\hline
\hline\\
\end{tabular}
\begin{minipage}{12cm}
\emph{Notes}: Cases I-IV denote ($p, ~(M_{\rm disk} + M_{\rm outflow}) / M_{\odot}$) = (0.3, 0.2), (0.8, 0.2), (0.3, 0.5), and (0.8, 0.5), respectively. $M_{\rm outflow}$ is the mass of disk outflows, and $\bar{P}_{r}$, $\bar{L}_{\rm bol}$, $\bar{L}_{\rm{\nu}}$, and $\bar{P}_{\rm GW}$ are the mean power injected by the $r$-process in the total outflows, the mean bolometric luminosity of the total outflows, the mean neutrino luminosity above (or below) the disk, and the mean GW power, respectively.
\end{minipage}
\end{center}
\label{table}
\end{table*}

The total GW energy can be calculated by
\begin{eqnarray}
E_{\rm{GW}}=\frac{\beta G}{9c^{5}}\int_{-\infty }^{\infty }L_{\rm{\nu}}(t)^{2}dt
\end{eqnarray}
with $\beta \sim 0.47039$.

The GW energy spectrum can be described by
\begin{eqnarray}
\frac{dE_{\rm GW}(f)}{df} =\frac{2 \beta G}{9 c^5} |\tilde{L}_{\nu} (f)|^2.
\end{eqnarray}

To determine the GW detectability of NDAFs, the characteristic strain of GWs is written as \citep[e.g.,][]{Flanagan1998}
\begin{eqnarray}
h_c (f)=\sqrt{\frac{2}{\pi ^2} \frac{G}{c^3} \frac{1}{D_{L}^2} \frac{dE_{\rm GW}(f)}{df}}.
\end{eqnarray}

After we obtained the correlation between $h_{\rm c} (f)$ and $f$ for a single pulse and multiple pulses from SGRBs, we presented the signal-to-noise ratios (S/Ns) of the corresponding filter for GWs. The optimal S/N is
\begin{eqnarray}
{\rm{S/N^{2}}}=\int_{0}^{\infty}d(\ln{f})\frac{h_{\rm c}(\emph{f})^{2}}{h_{\rm n}(\emph{f})^{2}},
\end{eqnarray}
where $h_{\rm n}(f) = [5f S_{\rm n}(f)]^{1/2}$ is the noise amplitude and $S_{\rm n}(f)$ is the spectral density of strain noise when the frequency of the detector is $f$.

\subsection{Results}

Figure~\ref{fig:GW1} shows $h_{\infty}/T$ as a function of the viewing angle $\theta_{\rm obs}$ at a distance of $10~\rm kpc$. It should be noted that there is almost no GW emission in the polar direction, whereas the GW amplitudes increase with increasing viewing angle.

Figure~\ref{fig:GW2} shows the characteristic strains of GWs from NDAFs with different outflow strengths in BNS/BH-NS merger scenarios at a distance of $10~\rm kpc$ for four cases of SGRBs with a single pulse (green solid lines in four figures, $T=0.3, 1, 0.1,$ and $1~\rm s$, respectively) and the multiple pulses (red solid lines in four figures, $N' = 100$, $\delta t = 0.001$ s, and $T=0.3, 1, 0.1,$ and $1~\rm s$, respectively). Moreover, for the cases of NDAFs without outflows, the green and red dashed lines represent the cases of SGRBs with a single pulse ($T=0.1$ s) and the multiple pulses ($N' = 100$, $\delta t = 0.001$ s, and $T=0.1$ s), respectively. The durations are determined by the accretion timescales. The black solid lines represent the sensitivity curves (the noise amplitudes $h_{n}$) of the following detectors, i.e., aLIGO, ET, CE, LISA, Taiji, TianQin, DECIGO/BBO, and ultimate-DECIGO.

We can notice that the typical GW frequencies are in the ranges of $\sim 1-1000$ and $\sim 10-1000$ Hz for the cases of NDAFs with and without outflows, respectively, which is determined by the GRB variabilities \citep[$f\sim 1/2\delta t$, see e.g.,][]{Liu2017b}. The GW waveforms induced by a single pulse and the multiple pulses are very different also caused by the distinction on the variabilities of the neutrino emission. The GW strains increase with decreasing outflow strengths, which corresponds increasing neutrino luminosities. Therefore, the GW strains reach the largest values when the NDAFs has no outflows. The mean GW power, $\bar{P}_{\rm GW}=E_{\rm{GW}}/T$, is shown in Table 1 for four cases. Indeed, the GW emission from NDAFs is originated from the anisotropic neutrino emission. One can note that GWs from NDAFs with and without outflows can be detected by ET, CE, DECIGO/BBO, and ultimate-DECIGO in the detectable frequency of $\sim 1-200~\rm Hz$ and $\sim 10-200~\rm Hz$ at $10~\rm kpc$ for SGRBs with a single pulse or the multiple pulses, respectively. The properties of this GWs are similar to those of GWs originated from a precessing jet launched by a warped NDAF \citep[e.g.,][]{Sun2012,Liu2017b}.

\section{Summary}

In a multimessenger study of BNS/BH-NS mergers, observed signals such as GRBs, kilonova, neutrinos, and GWs can be utilized to investigate the physics of BNS/BH-NS mergers. These multimessenger signals might mingle or even be dominated by the contributions of BH-NDAF systems. In this paper, we modeled NDAFs with outflows to study anisotropic DOD kilonovae, MeV neutrinos, and GWs in BNS/BH-NS merger scenarios. For DOD kilonovae, we find that the mass distributions do not significantly affect the anisotropy of the kilonova luminosity, but the velocity distribution influences the location of the radius of the photosphere, which in turn affects the luminosity. The anisotropy of DOD kilonovae appears in the J, u, and r bands but not in the K band. Then, we study the electron antineutrino energy spectra from NDAFs with and without outflows for different viewing angles. The general relativity has an important influence on the high-energy ($\gtrsim 10~\rm MeV$) neutrinos fluxes leading to anisotropy, but low-energy ($\lesssim 10~\rm MeV$) neutrinos fluxes are almost isotropic. MeV neutrinos from NDAFs in the Local Group might be detectable. Furthermore, we present the GW emission arising from anisotropic neutrino radiation from merger NDAFs with and without outflows. The GW stains are totally determined by the neutrino luminosities, and they increase with increasing viewing angles. The GW signals from NDAFs with outflows could be detected within the frequency of $1-200~\rm Hz$ at a distance of $10~\rm kpc$ by ET, CE, DECIGO/BBO, and ultimate-DECIGO.

As shown in Figures 3 and 4, it is difficult to constrain the viewing angles just according to the individual observations of the DOD kilonovae or MeV neutrinos, but the GW detections of the merger events can provide the morphology information of the mergers and the following BH hyperacctrion. Future joint multimessenger observations could provide effective insights into the central BH-NDAF systems created by BNS/BH-NS mergers at least in the Local Group and can be used to determine the nature of BH hyperaccretion models.

\acknowledgments
We thank Da-Bin Lin, Jin-Ping Zhu, Jun-Hui Liu, Jia Ren, and Zhen-Yu Zhu for helpful discussions. This work was supported by the National Natural Science Foundation of China under grants 11822304, 12173031, and 11773053.

\end{document}